\documentclass[10pt]{iopart}
\usepackage{amsfonts}

\begin{document}

\title{A homogeneous brane-world universe}
\author{L\'{a}szl\'{o} \'{A} Gergely \\
Institute of Cosmology and Gravitation, University of Portsmouth, Portsmouth
PO1 2EG, UK and\\
Astronomical Observatory and Department of Experimental Physics, University
of Szeged, Szeged 6720, D\'{o}m t\'{e}r 9, Hungary}

\begin{abstract}
A homogeneous, Kantowski-Sachs type, bouncing brane-world universe is
presented. The bulk has a \textit{positive} cosmological constant and the
Killing algebra $so(1,3)\oplus so(3)$. The totality of the source terms of
the effective Einstein equation combine to a solid with different radial and
tangential pressures.
\end{abstract}

\section{Introduction}

The 5-dimensional Einstein equation imposed on a bulk containing a single
brane located at $y=0$ (generalized Randall-Sundrum type II model \cite{RS}%
), with the energy-momentum tensor $T_{AB}$ (obeying $T_{AB}n^{B}=0$, where $%
n^{A}=\delta ^{A}{}_{y}$ is the unit normal to the brane) is 
\begin{equation}
\widetilde{G}_{AB}=\widetilde{\kappa }^{2}\left[ -\widetilde{\Lambda }%
\widetilde{g}_{AB}+\delta (y)\left\{ -\lambda g_{AB}+T_{AB}\right\} \right]
\,.  \label{1}
\end{equation}%
Here $\widetilde{\kappa }^{2}=8\pi /\widetilde{M}_{\mathrm{p}}^{3}$ is the
5-dimensional coupling constant ($\widetilde{M}_{\mathrm{p}}$ being the
5-dimensional Planck mass) and $\lambda $ is the brane tension. Due to the
Gauss-Codazzi equations and the junction conditions across the $Z_{2}$%
-symmetrically embedded brane, the effective Einstein equation on the
brane~emerges \cite{SMS}: 
\begin{equation}
G_{\mu \nu }=-\Lambda g_{\mu \nu }+\kappa ^{2}T_{\mu \nu }+\widetilde{\kappa 
}^{4}S_{\mu \nu }-\mathcal{E}_{\mu \nu }\,,  \label{2}
\end{equation}%
where $\kappa ^{2}=8\pi /M_{\mathrm{p}}^{2}$ is the 4-dimensional coupling
constant and $g_{AB}=\widetilde{g}_{AB}-n_{A}n_{B}$ is the induced metric.
The 5-dimensional and 4-dimensional energy scales and cosmological constants
are related to each other and the brane tension via 
\begin{eqnarray}
\widetilde{\kappa }^{4}\lambda &=&6{\kappa ^{2}}\,,  \nonumber \\
2\Lambda &=&\widetilde{\kappa }^{2}\widetilde{\Lambda }+{\kappa ^{2}}\lambda
\,.  \label{3}
\end{eqnarray}%
Here $S_{\mu \nu }$ are local quadratic energy-momentum corrections, given
by 
\begin{equation}
S_{\mu \nu }={{\frac{1}{12}}}T_{\alpha }{}^{\alpha }T_{\mu \nu }-{{\frac{1}{4%
}}}T_{\mu \alpha }T^{\alpha }{}_{\nu }+{{\frac{1}{24}}}g_{\mu \nu }\left[
3T_{\alpha \beta }T^{\alpha \beta }-\left( T_{\alpha }{}^{\alpha }\right)
^{2}\right] \,.  \label{4}
\end{equation}%
and $\mathcal{E}_{\mu \nu }$ is the $y\rightarrow 0$ limit of the $\mathcal{E%
}_{AB}=\widetilde{C}_{ACBD}n^{C}n^{D}$ projection of the bulk Weyl tensor.

Various solutions of this new scenario have already been considered, the
most recent of them being the G\"{o}del brane \cite{BTs}. Due to the
cosmological implications, likely to be experimentally tested, the Friedmann
branes were widely employed \cite{Maartens}, \cite{BDEL}, \cite{Decomp}. The
results of \cite{MSM} and \cite{BCG} suggested that all bulk solutions with
Friedmann branes should be 5-dimensional Schwarzschild-anti de Sitter
(SAdS). However a class of brane-world solutions generalizing the Einstein
static universe was found \cite{EinBrane}, which disobey this rule. The
simplest of them, the generalized Einstein static brane with curvature index 
$\epsilon =0$ is embedded in a \textit{flat} bulk in \cite{EinBrane}. The
solutions given in \cite{EinBrane} bear in common a negative energy density $%
\rho ^{E}=-\lambda $ of the perfect fluid source, however the effective
energy density (computed with respect to the unit tangent vector along the
fluid flow lines) from all contributions to the source (including the
4-dimensional cosmological constant) is $\kappa ^{2}\rho _{eff}^{E}=%
\widetilde{G}_{AB}u^{A}u^{B}=G_{\mu \nu }u^{\mu }u^{\nu }=\tilde{\Lambda}%
\widetilde{\kappa }^{2}$, positive for any bulk cosmological constant $%
\widetilde{\Lambda }>0$.

In this paper we present (among other metrics with peculiar signatures) a
related homogeneous brane-word solution, obtained by applying the complex
transformation 
\begin{equation}
t\rightarrow i\chi \ ,\qquad \chi \rightarrow it\ ,  \label{transf}
\end{equation}%
to the solutions found in \cite{EinBrane}. From a 4-dimensional point of
view the brane in this solution is a solid with spherical and homogeneous
symmetries and constant tensions.

The notation follows closely that of Ref. \cite{EinBrane}.

\section{Homogeneous universe}

As in \cite{EinBrane} we introduce the notation 
\begin{equation}
\widetilde{\kappa }^{2}\widetilde{\Lambda }=3\epsilon \Gamma ^{2}\,\,,\qquad
\Gamma >0\,\,.  \label{Ein5}
\end{equation}%
Here $\epsilon =0,\pm 1$ carries the sign of the bulk cosmological constant.
The one-parameter class of solutions found in \cite{EinBrane} is: 
\begin{equation}
\Gamma ^{2}d\,\widetilde{s}_{E}\,^{2}=-F^{2}(y;\epsilon )dt^{2}+d\chi ^{2}+%
\mathcal{H}_{E}^{2}\left( \chi ;\epsilon \right) \left( d\theta ^{2}+\sin
^{2}\theta d\varphi ^{2}\right) +dy^{2}\,,
\end{equation}%
with the metric functions 
\begin{equation}
F(y;\epsilon )=\left\{ 
\begin{array}{cc}
A\cos \left( \sqrt{2}\ y\right) +B\sin \left( \sqrt{2}\ y\right) \  & 
,\qquad \epsilon =1\,, \\ 
A+\sqrt{2}\ By & ,\qquad \epsilon =0\,, \\ 
A\cosh \left( \sqrt{2}\ y\right) +B\sinh \left( \sqrt{2}\ y\right) & \
,\qquad \epsilon =-1\,.%
\end{array}%
\right.  \label{F}
\end{equation}%
and 
\begin{equation}
\mathcal{H}_{E}(\chi ;\epsilon )=\left\{ 
\begin{array}{cc}
\cos \ \chi \  & ,\qquad \epsilon =1\,, \\ 
\chi & ,\qquad \epsilon =0\,, \\ 
\sinh \ \chi & \ ,\qquad \epsilon =-1\,.%
\end{array}%
\right. \   \label{H}
\end{equation}%
(For $\epsilon =1$ we have changed the function $\sin $ into $\cos $ . This
being a simple translation in the coordinate $\chi $, it does not change the
solution. This new form of the metric is required in order to obtain the new
solutions presented below.)

By the complex transformation (\ref{transf}) the line element becomes 
\begin{equation}
\Gamma ^{2}d\,\widetilde{s}\,^{2}=-dt^{2}+F^{2}(y;\epsilon )d\chi ^{2}+%
\mathcal{H}^{2}\left( t;\epsilon \right) \left( d\theta ^{2}+\sin ^{2}\theta
d\varphi ^{2}\right) +dy^{2}\ ,  \label{metric5ep}
\end{equation}%
with the metric function $\mathcal{H}(t;\epsilon )$ given by 
\begin{equation}
\mathcal{H}(t;\epsilon )=\left\{ 
\begin{array}{cc}
\cosh \ t & ,\qquad \epsilon =1\,, \\ 
it & ,\qquad \epsilon =0\,, \\ 
i\sin \ t & \ ,\qquad \epsilon =-1\,.%
\end{array}%
\right. \ \ .  \label{Hnew}
\end{equation}%
Alternatively, we can define $\mathcal{H}$ by relations very similar to the
ones given in \cite{EinBrane}: 
\begin{equation}
\left( \partial _{t}\mathcal{H}\right) ^{2}=\epsilon \mathcal{H}^{2}-1\
,\qquad \qquad \partial _{\chi }^{2}\mathcal{H}=\epsilon \mathcal{H}\,.
\end{equation}%
The new metrics solve the Einstein equations (\ref{1}) in the bulk. In the
cases $\epsilon =0,-1$ the metric has the signature $\left( -+--+\right) $
and we do not study further these solutions.

The $\epsilon =1$ case gives the new one-parameter solution:%
\begin{eqnarray}
\Gamma ^{2}d\,\widetilde{s}\,^{2} &=&-dt^{2}+\left[ A\cos \left( \sqrt{2}\
y\right) +B\sin \left( \sqrt{2}\ y\right) \right] ^{2}d\chi ^{2}  \nonumber
\\
&&+\cosh ^{2}t\ \left( d\theta ^{2}+\sin ^{2}\theta d\varphi ^{2}\right)
+dy^{2}\ .  \label{metric5}
\end{eqnarray}%
The symmetries of this metric form the Killing algebra $so(1,3)\oplus so(3)$%
, as shown in the Appendix.

The projected part of the bulk Weyl tensor characterizing the metric (\ref%
{metric5}) is

\begin{equation}
\mathcal{E}_{AB}=\tilde{C}_{ACBD}n^{C}n^{D}={{\frac{1}{2}}}\Gamma ^{2}\left(
u_{A}u_{B}+3e_{A}e_{B}-l_{AB}\right) \,,  \label{Weylproj}
\end{equation}%
where $u^{A}=\Gamma \delta ^{A}{}_{0}$ is the unit 4-velocity along $%
\partial /\partial t$, $e^{A}=\Gamma F^{-1}\delta ^{A}{}_{1}$ is the
homogeneous Killing vector of the metric~(\ref{metric5}) and $%
l_{AB}=g_{AB}+u_{A}u_{B}-e_{A}e_{B}$.

On the brane ($y=0$) the metric (\ref{metric5}) induces the 4-metric:

\begin{equation}
\Gamma ^{2}d\,s^{2}=-dt^{2}+A^{2}d\chi ^{2}+\cosh ^{2}t\ \left( d\theta
^{2}+\sin ^{2}\theta d\varphi ^{2}\right) \ .  \label{metric4}
\end{equation}%
As the continuity of the metric across the brane requires the same value of
the constant $A$ on both sides of the brane, the constant $A$ can be
absorbed into the coordinate$\ \chi $. The brane has spherical symmetry but
it is also homegeneous, as is the Kantowski-Sachs metric \cite{KS} of
ordinary 4-dimensional general relativity. Having the scale-factors $%
1/\Gamma $ and $\cosh t/\Gamma $, it is different from the Kantowski-Sachs
brane cosmologies presented in \cite{MOO}. The metric (\ref{metric4}) has
positive curvature $R=6\Gamma ^{2}$, therefore it is not contained in the
Bianchi classification \cite{BD}. However, as shown in the Appendix, it has
even more symmetries. The remaining three Killing symmetries are boost-like.
The complete algebra of the Killing vectors on the brane is $so(1,3)\oplus 
\mathbb{R}$. The scale-factor $\cosh t$ implies a bouncing character of this
brane-universe in the time-parameter $t$.

The extrinsic curvature of any hypersurface $y=$const. in the metric (\ref%
{metric5}) has only one nonvanishing component

\begin{equation}
K_{\chi \chi }={\frac{1}{\Gamma }}F\left( y\right) \partial _{y}F\left(
y\right) \,\ .
\end{equation}%
In consequence the jump in the extrinsic curvature across the brane (at $y=0$%
) is

\begin{equation}
\lbrack K_{\mu \nu }]={\sqrt{2}[B]}\Gamma \,e_{\mu }e_{\nu }\,.
\label{Kjump}
\end{equation}%
A ${Z}_{2}$-symmetric bulk solution is given by Eq.~(\ref{metric5}) with $y$
replaced by $|y|$. Then $[B]$ is replaced by $2B$ in Eq. (\ref{Kjump}).

The relation $[K_{\mu \nu }]=\widetilde{\kappa }\,^{2}\left( \left[ T_{\mu
\nu }+\frac{1}{3}\left( \lambda -T\right) g_{\mu \nu }\right] \right) $
gives the brane source 
\begin{eqnarray}
T_{\mu \nu } &=&\rho u_{\mu }u_{\nu }+\lambda e_{\mu }e_{\nu }-\rho l_{\mu
\nu }\ ,  \nonumber \\
\widetilde{\kappa }^{2}\left( \rho +\lambda \right) &=&2{\sqrt{2}B}\Gamma \ .
\label{enmom}
\end{eqnarray}%
The energy-momentum tensor describes a perfect fluid with negative energy
density $\rho =-\lambda $ only in the case $B=0$. This condition would imply
no jump in the extrinsic curvature, therefore the matter on the brane being
generated as a response to the brane tension $\lambda $. In the generic case
the energy-momentum tensor (\ref{enmom}) represents a generalized fluid. As $%
\lambda >0$, the possibility of a positive energy density is left open, in
contrast with the solutions presented in \cite{EinBrane}. Then the radial
principal pressure is positive, while the tangential ones are negative.

The quadratic source term (\ref{4}) gives 
\begin{equation}
\widetilde{\kappa }^{4}S_{\mu \nu }=\frac{\kappa ^{2}\lambda }{2}g_{\mu \nu
}-\kappa ^{2}T_{\mu \nu }\,,
\end{equation}%
and the totality of the source terms in the effective 4d Einstein equation
combine to 
\begin{equation}
G_{\mu \nu }=\Gamma ^{2}\left( u_{A}u_{B}-3e_{A}e_{B}-l_{AB}\right) \,.
\label{source}
\end{equation}%
(In the derivation we have employed Eqs. (\ref{3}) and (\ref{Ein5}) several
times.)

\section{Concluding remarks}

We have presented a homogeneous brane-world universe obtained from the
Einstein brane-universe by applying the complex transformation (\ref{transf}%
). This universe has a bouncing character. Its source given by Eq. (\ref%
{enmom}) is difficult to interpret in terms of ordinary matter. However,
from the classical general relativistic point of view, on the right hand
side of Eq. (\ref{source}) we have a generalized solid with energy density $%
\kappa ^{2}\rho _{eff}=\Gamma ^{2}$, radial tension $\kappa
^{2}p_{eff}^{rad}=-3\Gamma ^{2}$ and tangential tension $\kappa
^{2}p_{eff}^{\tan }=-\Gamma ^{2}$. This is one of the rare solutions of the
Einstein field equations with symmetry group $G_{7}$, from among which only
the Einstein static universe and some special plane waves were known
previously \cite{ExSol}.

It is worth to note, that the Einstein static brane presented in \cite%
{EinBrane}, with $\rho ^{E}=-\lambda <0$ and $\widetilde{\kappa }^{2}\left(
\rho ^{E}+p^{E}\right) =-2{\sqrt{2}B}\Gamma $, can be interpreted from the
general relativistic point of view as having effective energy density $%
\kappa ^{2}\rho _{eff}^{E}=3\epsilon \Gamma ^{2}$ and isotropic tension $%
\kappa ^{2}p_{eff}^{E}=-\epsilon \Gamma ^{2}$, a \textit{perfect} solid for $%
\epsilon =1$.

\ack

I am grateful to Roy Maartens for discussions. This work was supported by
the Hungarian E\"{o}tv\"{o}s Fellowship.

\appendix

\section{Killing vectors}

A set of independent Killing vectors for the metric (\ref{metric5ep}) in the
coordinates $x^{A}=(t,\chi ,\theta ,\varphi ,y)$ is given by

\begin{eqnarray}
K_{\mathbf{1}} &=&\left( 0,0,0,1,0\right) \ ,  \nonumber \\
K_{\mathbf{2}} &=&\left( 0,0,-\cos \varphi ,\cot \theta \sin \varphi
,0\right) \ ,  \nonumber \\
K_{\mathbf{3}} &=&\left( 0,0,\sin \varphi ,\cot \theta \cos \varphi
,0\right) \ ,  \nonumber \\
K_{\mathbf{4}} &=&\left( -\cos \theta ,0,\partial _{t}\left( \log \mathcal{H}%
\right) \sin \theta ,0,0\right) \ ,  \nonumber \\
K_{\mathbf{5}} &=&\left( \sin \theta \sin \varphi ,0,\partial _{t}\left(
\log \mathcal{H}\right) \cos \theta \sin \varphi ,\partial _{t}\left( \log 
\mathcal{H}\right) \frac{\cos \varphi }{\sin \theta },0\right) \!\ \!\!, 
\nonumber \\
K_{\mathbf{6}} &=&\left( \sin \theta \cos \varphi ,0,\partial _{t}\left(
\log \mathcal{H}\right) \cos \theta \cos \varphi ,\!-\!\partial _{t}\left(
\log \mathcal{H}\right) \frac{\sin \varphi }{\sin \theta },0\!\right) \ \!\!,
\nonumber \\
K_{\mathbf{7}} &=&\left( \alpha \beta \right) ^{-1}\left( 0,1,0,0,0\right) \
,  \nonumber \\
K_{\mathbf{8}} &=&\frac{\alpha ^{2}}{\sqrt{2}\beta }\left( 0,-\partial
_{y}\left( \log F\right) \sin \left( \beta \chi \right) ,0,0,\beta \cos
\left( \beta \chi \right) \right) \ ,  \nonumber \\
K_{\mathbf{9}} &=&-\frac{\alpha }{\sqrt{2}\beta }\left( 0,\partial
_{y}\left( \log F\right) \cos \left( \beta \chi \right) ,0,0,\beta \sin
\left( \beta \chi \right) \right) \ ,
\end{eqnarray}%
where we have introduced the notations $\alpha =sgn\left( \epsilon
A^{2}+B^{2}\right) $ and $\beta =\sqrt{2\mid \epsilon A^{2}+B^{2}\mid }.$
These Killing vectors obey 
\begin{eqnarray}
\left[ K_{\mathbf{i}},K_{\mathbf{j}}\right] &=&\varepsilon _{ijk}K_{\mathbf{k%
}}\ ,\qquad  \nonumber \\
\left[ K_{\mathbf{3+i}},K_{\mathbf{3+j}}\right] &=&-\varepsilon _{ijk}K_{%
\mathbf{k}}\ ,\qquad  \nonumber \\
\left[ K_{\mathbf{i}},K_{\mathbf{3+j}}\right] &=&\varepsilon _{ijk}K_{%
\mathbf{3+k}}\ ,  \nonumber \\
\left[ K_{\mathbf{6+i}},K_{\mathbf{6+j}}\right] &=&\varepsilon _{ijk}K_{%
\mathbf{6+k}}\ ,\qquad  \nonumber \\
\left[ K_{\mathbf{6+i}},K_{\mathbf{j}}\right] &=&0=\left[ K_{\mathbf{6+i}%
},K_{\mathbf{3+j}}\right] \ .\qquad \qquad  \label{KilAlg}
\end{eqnarray}%
Thus the Killing algebra is $so(1,3)\oplus so(3)$. From among the Killing
vectors $K_{\mathbf{1-7}}$ are confined to the $y=const$ sections. They span
the algebra $so(1,3)\oplus \mathbb{R}$.

The vectors $K_{\mathbf{1-7}}$ (without the fifth component) are the Killing
vectors for the brane metric (\ref{metric4}). Among them $K_{\mathbf{1-3}}$
and $K_{\mathbf{7}}$ are spacelike and they assure the homogeneity of the
constant time slices (they form the $so(3)\oplus \mathbb{R}$ algebra). The
causal character of the remaining three Killing vectors is given by%
\begin{eqnarray*}
\Gamma ^{2}g\left( K_{\mathbf{4}},K_{\mathbf{4}}\right) &=&-1+\sin
^{2}\theta \cosh ^{2}t\ , \\
\Gamma ^{2}g\left( K_{\mathbf{5}},K_{\mathbf{5}}\right) &=&-1+\left( 1-\sin
^{2}\theta \sin ^{2}\varphi \right) \cosh ^{2}t\ , \\
\Gamma ^{2}g\left( K_{\mathbf{6}},K_{\mathbf{6}}\right) &=&-1+\left( 1-\sin
^{2}\theta \cos ^{2}\varphi \right) \cosh ^{2}t\ .
\end{eqnarray*}%
For $t=0$ all three of them are time-like, while for other values of $t$ the
causal character depends on the actual values of $\theta $ and $\varphi $.
The region with all three Killing vectors spacelike increases with $\mid
t\mid .$


\section*{References}

\begin{thebibliography}{99}
\bibitem{RS} L Randall and R Sundrum 1999 \textit{Phys. Rev. Lett.} \textbf{%
83 }4690

\bibitem{SMS} T Shiromizu, K Maeda and M Sasaki 2000 \textit{Phys. Rev. D }%
\textbf{62} 024012

\bibitem{BTs} J D Barrow, C G Tsagas 2003 \textit{The G\"{o}del Brane}
gr-qc/0309030\textit{\ }

\bibitem{Maartens} R Maartens 2000 \textit{Phys. Rev.} D \textbf{62} 084023

\bibitem{BDEL} P Bin\'{e}truy, C Deffayet, U Ellwanger and D Langlois 2000 
\textit{Phys. Lett.} B \textbf{477} 285

\bibitem{Decomp} L \'{A} Gergely 2003 \textit{Generalized Friedmann Branes}
gr-qc/0308072 (see also references therein)

\bibitem{MSM} S Mukohyama, T Shiromizu and K Maeda 2000 \textit{Phys. Rev. D 
}\textbf{62\ }024028

\bibitem{BCG} P Bowcock, C Charmousis and R Gregory 2000 \textit{Class.
Quantum Grav. }\textbf{17} \ 4745

\bibitem{EinBrane} L \'{A} Gergely and R Maartens 2002 \textit{Class.
Quantum Grav. }\textbf{19} \ 213

\bibitem{KS} R Kantowski, R K Sachs 1966 \textit{J. Math. Phys. }443

\bibitem{MOO} A N Makarenko, V V Obukhov, K E Osetrin 2003 \textit{%
Kantowski-Sachs Brane Cosmology} gr-qc/0301124

\bibitem{BD} J D Barrow, M P D\c{a}browski 1997 \textit{Phys. Rev. D }%
\textbf{55 }630

\bibitem{ExSol} H Stephani, D Kramer, M MacCallum, C Hoensalaers and E Hertl
2003 \textit{Exact Solutions of Einstein's Field Equations, Second Edition},
Cambridge Univ. Press
\end{thebibliography}
\end{document}